\documentstyle[epsf,floats,tighten,preprint,prl,aps]{revtex}

\def\be{\begin{equation}}
\def\ee{\end{equation}}
\def\diag{\mathop{\rm diag}}

\begin{document}
\draft
\title{Superluminal travel requires negative energies}

\author{Ken D.\ Olum\footnote{Email address: {\tt kdo@alum.mit.edu}}}

\address{Institute of Cosmology \\
Department of Physics and Astronomy \\
Tufts University \\
Medford, MA 02155}

\date{August 1998}

\maketitle

\begin{abstract}
I investigate the relationship between faster-than-light travel and
weak-energy-condition violation, i.e., negative energy densities.  In
a general spacetime it is difficult to define faster-than-light
travel, and I give an example of a metric which appears to allow
superluminal travel, but in fact is just flat space.  To avoid such
difficulties, I propose a definition of superluminal travel which
requires that the path to be traveled reach a destination surface at
an earlier time than any neighboring path.  With this definition (and
assuming the generic condition) I prove that superluminal travel
requires weak-energy-condition violation.
\end{abstract}

\narrowtext

A longstanding question asks whether the metric of spacetime can be
manipulated to allow very rapid travel between spatially distant
points.  (I will call this ``superluminal'' or ``faster than light'' even
though, of course, I'm not proposing to go faster than a light signal
in the same metric).  If one allows arbitrary states of matter, one
can construct such spacetimes, as in the examples of
Alcubierre\cite{Alcu94} and Krasnikov\cite{Kras95,E&Ro97}.  However, these
spacetimes require negative energy densities\cite{E&Ro97,Pfen97b};
i.e., they violate the weak energy condition (WEC), which states that
$T_{\mu\nu} V^\mu V^\nu \geq 0$ for any timelike vector $V^\mu$.  The
question then is whether it is possible to have superluminal travel
without this violation.

To answer this question one must first specify what one means by
``superluminal travel.''  The general idea is that some modification of
the metric allows signals to propagate between two spacetime points
that otherwise would be causally disconnected.  However, it may not
always be easy to distinguish such superluminal travel from a
situation in which the supposedly distant object has been brought
nearby, so that travel at ordinary speeds allows one to reach it in a
short time.

\begin{figure}
\begin{center}
\leavevmode\epsfbox{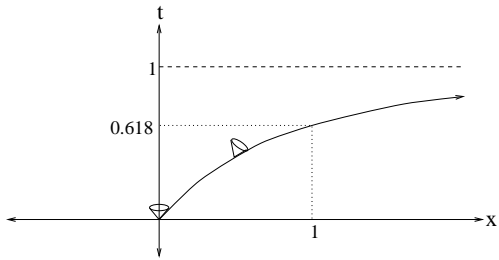}
\end{center}
\caption{A null geodesic in the metric of Eq.\ (\ref{eqn:metric}).  It
appears that one can reach arbitrary distances before $t = 1$.} 
\label{fig:geodesic}
\end{figure}

As a concrete example consider a spacetime with metric
\be\label{eqn:metric}
ds^2 = (-1 + 4t^2x^2) dt^2-4tx (1-t^2) dxdt + (1-t^2)^2 dx^2
\ee
in the region $-1 < t < 1$.
Null rays in this metric have
\be
{dx\over dt} ={\pm 1 + 2tx\over 1-t^2}\,,
\ee
and, for example, a right-going null geodesic from the origin has
$x = t/(1-t^2)$ as shown in Fig.\ \ref{fig:geodesic}.
It would appear that this metric allows superluminal travel.  Starting
from the origin one can reach points at arbitrarily large $x$ in time
$t<1$.  If the earth were fixed at $x=0$ and a distant star at $x=1$,
we could travel from the earth at $t = 0$ to the star in time $(1
+\sqrt{5})/2\approx 0.618$.

However, this metric has nothing to do with superluminal travel.  It
is just flat space with an odd choice of coordinates: if we let $x'= x
(1-t^2)$ the metric becomes $ds^2 = -dt^2 + dx'^2$.  The star which is
``fixed'' at $x = 1$ is in fact traveling on a path which brings it
closer to the earth.  The motion of the destination, rather than any
superluminal travel, is what reduces the time to reach the star.

The point of this example is that just examining a metric may not
easily reveal whether it would be reasonable to regard the spacetime
as containing superluminal travel.  One must have some idea how to
distinguish bringing a place closer from establishing an arrangement
which allows one to travel there more quickly.

\begin{figure}
\begin{center}
\leavevmode\epsfbox{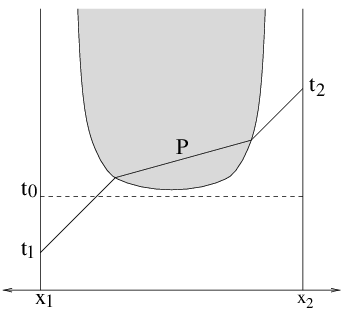}
\end{center}
\caption{Superluminal travel is produced by modifying the shaded region of
Minkowski space.  The modification is localized between
$x_1$ and $x_2$ and after $t_0$.  Because of this modification, there
is a causal path $P$ connecting $(t_1, x_1)$ to $(t_2, x_2)$, even
though $x_2-x_1 > t_2-t_1$.} 
\label{fig:ftl-minkowski}
\end{figure}

In some simple cases, however, the spacetime is flat, except for a
localized region not including the points between which one wishes to
travel.  Then there is no question about the distance between the two
points, because they lie in a single region of Minkowski space.  The
Alcubierre bubble\cite{Alcu94} and the Krasnikov
tube\cite{Kras95,E&Ro97} are of this type if one imagines the tube to
be finite in length or the bubble to exist for a finite time.  A simple
example of this sort is shown in Fig.\ \ref{fig:ftl-minkowski}.
The flat metric has been modified in such a way that there is a causal
path $P$ from $(t_1, x_1)$ to $(t_2, x_2)$ even though $x_2-x_1 >
t_2-t_1$.  Since there is a connected region of Minkowski space which
includes $(t_1, x_1)$ and $(t_2, x_2)$, it is well-defined to say that
the interval between these points would be spacelike without the
modification to the metric.

In this simple case we can show that WEC must be violated, using the
existing theorems\cite{tip76,tip77,cpc} that prohibit closed timelike
curves.  Let $S$ be a spacetime that is flat except for a region with
$t > t_0$, $x \in [x_1, x_2]$, $y \in [y_1, y_2]$, and $z \in [z_1,
z_2]$, and suppose there is a causal path $P$ that connects points
$(t_1, x_1, y_0, z_0)$ and $(t_2, x_2, y_0, z_0)$ with $t_2-t_1 <
x_2-x_1$.  Suppose also that $S$ contains no singularities and that
the modified region of $S$ obeys the generic condition\cite{Ha&El},
i.e., each null geodesic that passes through that region contains a
point where $K_{[a} R_{b]cd [e} K_{f]} K^cK^d\neq 0$, where $K$ is the
tangent vector to the geodesic.  Let $\Delta t = t_2-t_1$.  Consider a
new spacetime $S'$ which consists of the portion of $S$ between $x_1$
and $x_2$ with the same metric as $S$, and with points $(t, x_1, y,
z)$ and $(t +\Delta t, x_2, y, z)$ identified for each $t$, $y$, and
$z$.  In $S'$, the path $P$ is a closed causal curve.  However, causal
paths that travel only through the flat part of $S'$ cannot be closed,
because $\Delta t < x_2-x_1$.  In particular no point with
$t<t_0-\Delta t$ can be on a closed causal path.  So there is a Cauchy
horizon in $S'$ in the future of the surface $t = t_0 - \Delta t$ and
in the past of (or at) the path $P$.  If $S$ has no singularities,
than $S'$ will not have any either.  Thus by Tipler's and Hawking's
theorems\cite{tip76,tip77,cpc}, WEC must be violated somewhere in
$S'$.  Since WEC is a local condition, it must also be violated at the
corresponding point of $S$.

\begin{figure}
\begin{center}
\leavevmode\epsfbox{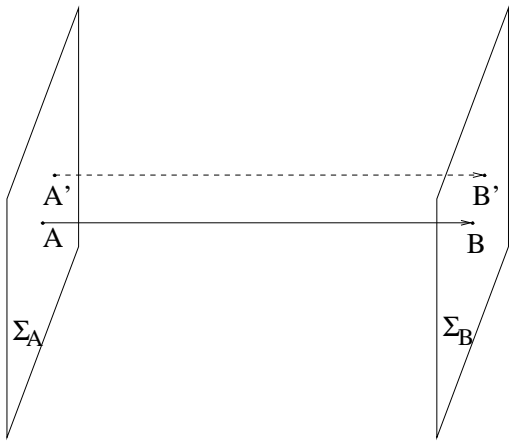}
\end{center}
\caption{A superluminal travel arrangement.  The metric has been so
arranged so that a causal path (solid line) exists between $A$ and $B$
but there are no other causal paths (such a possibility is shown dashed) that
connect the 2-surfaces $\Sigma_A$ and $\Sigma_B$.} 
\label{fig:basic}
\end{figure}

In a general spacetime we need a definition of
superluminal travel.  Here I propose the following idea: a superluminal
travel arrangement should have some path along which it functions
best.  A signal propagating along this best path should travel further
than a signal on any nearby path in the same amount of (externally
defined) time.  To formalize this we construct small spacelike
2-surfaces around the origin and destination points and say that while
the destination is reachable from the origin, no other point of the
destination surface is reachable from any point of the origin surface.
See Fig.\ \ref{fig:basic}.
Of course this would be trivial if, for example, the destination
surface were curved in such a way that the destination were merely the
closest point on its surface to the origin.  To avoid this problem we
require that the origin (destination) surface be composed of a
one-parameter family of spacelike geodesics through the origin
(destination) point.  Formally, we say that a causal path $P$ is
superluminal from $A$ to $B$ only if it satisfies

\vskip 10pt
\noindent
{\bf Condition 1}
{\it
There exist 2-surfaces $\Sigma_A$ around $A$ and
$\Sigma_B$ around $B$ such that (i) if $p\in\Sigma_A$ then a spacelike
geodesic lying in $\Sigma_A$ connects $A$ to $p$, and similarly
for $\Sigma_B$, and (ii) if $p\in\Sigma_A$ and $q\in\Sigma_B$
then $q$ is in the causal future of $p$ only if $p = A$ and $q = B$.
}
\vskip 10pt

This condition might not be sufficient for what one would call
superluminal travel, because it is possible that while $P$ arrives
earlier than any nearby path, it is still slower than a path some
larger distance away.  In this case, we would not want to say that $P$
was superluminal.

Suppose that there is a path $P$ satisfying the above condition, and
suppose also that the generic condition\cite{Ha&El} holds on $P$.  The
generic condition holds whenever there is any normal matter or any
transverse tidal force anywhere on $P$.  With these assumptions, we
will show that WEC must be violated at some point of $P$.

First we note that $P$ must be a null geodesic.  If $P$ is not a
geodesic it can be varied to make a timelike path from $A$ to $B$.  If
$P$ is timelike anywhere, then it can be varied to make a timelike
path from $A$ to points of $\Sigma_B$ other than $B$.

Let $K$ be the tangent vector to the geodesic $P$.  The vector $K$
must be normal to the surface $\Sigma_A$.  Otherwise there would be
points on $\Sigma_A$ in the past of points on $P$.  Similarly, $K$
must be normal to $\Sigma_B$.

Now define a congruence of null geodesics with affine parameter $v$,
normal to $\Sigma_A$, and extend $K$ to be the tangent vector at each
point of the congruence.

Could there be some point $x\in P$ that is conjugate to the surface
$\Sigma_A$?  If $x$ were an interior point of $P$ then it would be
possible to deform $P$ into a timelike path.  If $x = B$ then different
geodesics of the congruence would all end at $B$ or points very near
to $B$.  These geodesics would have different tangent vectors, which
could not all be normal to $\Sigma_B$.  Thus no point on $P$ is
conjugate to $\Sigma_A$.

Now we look at $\hat\theta$, the expansion of the geodesic congruence.
It is given by $\hat\theta = {K ^ m}_{;m}$, where $m$ runs over two
orthogonal directions normal to $K$.  (All choices of such directions
give the same $\hat\theta$.)  At $A$ we use directions that lie in
$\Sigma_A$ and at $B$ we use directions that lie in $\Sigma_B$.
Since $\Sigma_A$ is extrinsically flat at $A$, the geodesics are
initially parallel, so $\hat\theta = 0$ at $A$.  The evolution of
$\hat\theta$ is given by the Raychaudhuri equation for null geodesics,
\be\label{eqn:ray}
{d\hat\theta\over dv} = -R_{ab} K ^ aK ^ b + 2\hat\omega ^ 2-2\sigma ^
2-{1\over 2}\hat\theta ^ 2
\ee
where $\hat\omega$ is the vorticity, which vanishes here, $\hat\sigma$
is the shear, and $R_{ab}$ is the Ricci curvature tensor.  Since there
are no conjugate points, $\hat\theta$ is well-defined all along $P$.
If the weak energy condition is satisfied, then $R_{ab} K^aK^b\ge 0$,
so ${d\hat\theta/dv\leq 0}$.  From the generic condition, $\hat\sigma$
cannot vanish everywhere, thus WEC implies
\be\label{eqn:converging}
\hat\theta <0
\ee
at $B$.  If we can show
that instead $\hat\theta\geq 0$ at $B$, then WEC must be violated on
$P$.

First we establish a basis for vectors at $B$.  Let $E_1$ and $E_2$ be
orthonormal vectors tangent to $\Sigma_B$ at $B$.  Let $E_3$ be a unit
spacelike vector orthonormal to $E_1$ and $E_2$ with $g (K, E_3) > 0$.
Let $E_4$ be the unit future-directed timelike vector orthogonal to
$E_1$, $E_2$, and $E_3$.  Using these vectors establish (Riemannian)
normal coordinates near $B$.  The space $\Sigma_B$ consists of the
points with $t = z = 0$.

\begin{figure}
\begin{center}
\leavevmode\epsfbox{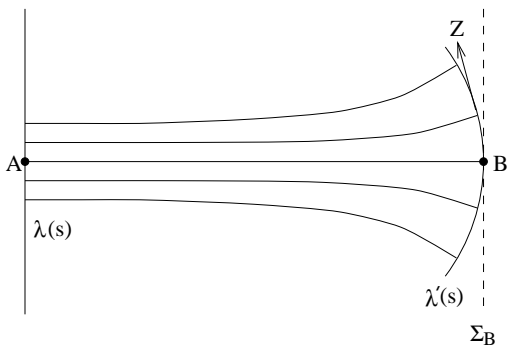}
\end{center}
\caption{Congruence of null geodesics from $\lambda (s)$ followed into
the future until they reach points near $B$ with $t = 0$ at a curve
$\lambda' (s)$ with tangent vector $Z$.  At points near $B$, $\lambda'
(s)$ must have negative $z$ coordinate.} 
\label{fig:lambda}
\end{figure}

Let $\lambda (s)$ be a smooth curve on $\Sigma_A$ with $\lambda (0) =
A$.  Let $\lambda (s, v) $ be the point an affine distance $v$ along
the null geodesic from $\lambda (s)$.  Eventually each geodesic will pass
near $B$ and will cross the hypersurface with $t = 0$.  Call this
crossing point $\lambda' (s)$ and adjust the length of the vectors $K$
on $\Sigma_A$ so that $\lambda(s, 1) =\lambda'(s)$.  See Fig.\
\ref{fig:lambda}.

The $z$ coordinate of $\lambda' (s)$ must be negative.  Otherwise,
points on $\Sigma_B$ ($z = t = 0$) would be the future of points of
the geodesics from $\Sigma_A$.

Let $Z$ be the tangent vector to $\lambda (s, v)$ in the $s$
direction.  By construction, $K ^ aZ_a = 0$ on $\Sigma_A$.  This
product is constant along each geodesic\cite{Ha&El}, so $K ^ aZ_a = 0$
everywhere.  If we follow along $\lambda' (s)$ from $B$ we see that
\be\label{eqn:perpendicular}
0 = {d\over ds} (K^aZ_a) = (K ^ aZ_a)_{; b} Z ^ b = {K ^ a} _ {;
b} Z_aZ ^ b + K ^ aZ_{a; b} Z ^ b\,.
\ee
The only non-vanishing components of $K$ are $K ^ 3$ and $K ^ 4$.
Since $\lambda '(s)$ lies in the $t = 0$ hypersurface, $Z ^ 4 = 0$
everywhere, so only $a = 3$ contributes to $K ^ a Z_{a; b}$ at $B$.  Thus
from Eq.\ (\ref{eqn:perpendicular})
\be
{K ^ a} _ {; b} Z_aZ ^ b = -K ^ 3Z_{3; b} Z ^ b\,.
\ee
At $B$, $Z_3 = 0$.  We must also have $Z_{3; b} Z ^ b\le 0$ because
otherwise ${\lambda'} ^ 3$ would become positive.  By construction, $K ^ 3 >
0$, so $K^3Z_{3; b} Z^b\le 0$ and
\be
{K ^ a} _ {; b} Z_aZ ^ b\ge 0\,.
\ee
The congruence of geodesics provides a map from tangent vectors to
$\lambda (s)$ at $A$ to tangent vectors to $\lambda' (s)$ at $B$.
Since there are no conjugate points, this map is non-singular and can
be inverted.  Thus we can find choices of $\lambda (s)$ that make $Z =
E_1$ or $Z = E_2$.  Then we find that ${K ^ 1}_{; 1}\ge 0$ and ${K ^
2}_{; 2}\ge 0$ and so
\be
\hat\theta = {K ^ m}_{; m}\ge 0
\ee
in contradiction to Eq. (\ref{eqn:converging}).

Thus we see that any spacetime that admits superluminal travel on
some path $P$ (and thus, according to our definition, that satisfies
Condition 1) and that satisfies the generic condition on $P$, must
also violate the weak energy condition at some point of $P$.

One can compare this theorem with those of Tipler\cite{tip76,tip77}
and Hawking\cite{cpc} that we used earlier.  These theorems rule out
the construction of closed timelike curves (CTC's) from a compact
region unless there is WEC violation or a singularity on the boundary
of the causality violating region.  The present theorem rules out the
existence, rather than construction, of superluminal travel, unless
there is weak energy condition violation.  Spacetime singularities do
not provide an alternative (other than by making the purported path
not actually reach the destination), and the WEC violation must occur
along the path to be traveled.

This raises the question of whether the present theorem can be
extended to rule out more time machines than the theorems of Tipler
and Hawking do.  However, this extension is not easily accomplished.
Inside a CTC-containing region, each point will be in the future of
each other point.  Thus one cannot construct surfaces $\Sigma_A$ and
$\Sigma_B$ with the required properties.  Even if one puts the points
$A$ and $B$ on the Cauchy horizon, it is still not possible to
construct spacelike 2-surfaces that do not intersect the
CTC-containing region.

Does this theorem mean that superluminal travel is impossible?  No,
because the weak energy condition is not obeyed by systems of quantum
fields.  The best example is the Casimir effect, and in fact, the Casimir
effect does provide an example which satisfies condition 1.

\begin{figure}
\begin{center}
\leavevmode\epsfbox{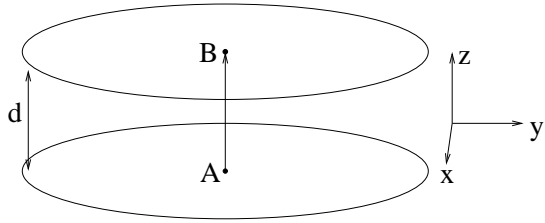}
\end{center}
\caption{Circular conducting plates give
rise to a negative pressure and energy density, and a consequent
advancement of the time of arrival of a null ray from $A$ to $B$.} 
\label{fig:Casimir}
\end{figure}

Consider the system shown in Fig.\ \ref{fig:Casimir}.
The quantum expectation value of the electromagnetic stress-energy
tensor between the plates is
\be
T_{ab} ={\pi^ 2\over 720d^4}\diag(-1, 1, 1, -3)\,.
\ee
For a geodesic traveling in the $z$ direction, we find
\be
R_{ab} K ^ aK ^ b = -{2\pi ^ 3\over 45d ^ 4}\,.
\ee
Now let $\Sigma_A$ be the lower plate and $\Sigma_B$ be the upper
plate, and we can go through the argument above in reverse.  We start
with $\hat\theta = 0$ as before, and now $\hat\sigma = 0$ by symmetry.
As before, $\hat\omega = 0$, so the Raychaudhuri equation
(\ref{eqn:ray}) gives
\be
{d\hat\theta\over dv} = -R_{ab} K ^ aK ^ b > 0
\ee
so the geodesics around $P$ are defocused.  Thus the geodesic $P$
travels further in the $z$ direction by the same $t$ than neighboring
geodesics, and condition 1 is satisfied.

It is not clear whether this phenomenon is sufficient to provide a
system of superluminal travel.  The discussion above is not complete,
because it does not account for the mass of the plates or of the
supporting structure required to keep them apart against the tension
in the field.  A long, straight path traveling through the center of
the plates arrives earlier than nearby paths, but it is possible that a
path that avoids the system of plates entirely might arrive still
earlier.

I would like to thank Arvind Borde, Allen Everett, Larry
Ford, Michael Pfenning, and Tom Roman for helpful conversations. This
work was supported in part by funding provided by the National Science
Foundation.

Note added in proof: While this paper was in press, I learned of
unpublished work by R. Penrose, R. D. Sorkin and E. Woolgar
(gr-qc/9301015) which also discusses the connection between WEC
violation and geodesic advancement.

\end{document}